\documentclass[twocolumn,showpacs,preprintnumbers,amsmath,amssymb]{revtex4}

\usepackage{graphicx}
\usepackage{dcolumn}
\usepackage{bm}
\usepackage{color}

\begin{document}

\preprint{APS/123-QED}

\title{Strong Suppression of Coherence Effect and Appearance of Pseudogap in Layered Nitride Superconductor Li$_x$ZrNCl: \\
$^{91}$Zr- and $^{15}$N- NMR Studies}

\author{Hisashi Kotegawa$^{1}$, Satoru Oshiro$^1$, Yuki Shimizu$^1$, Hideki Tou$^1$, Yuichi Kasahara,$^2$ Tsukasa Kishiume,$^3$ Yasujiro Taguchi,$^4$ and Yoshihiro Iwasa$^{2,4}$}

\affiliation{
$^1$Department of Physics, Kobe University, Kobe 657-8501, Japan \\
$^2$Quantum-Phase Electronics Center, School of Engineering, University of Tokyo, Tokyo 113-8656, Japan \\
$^3$Institute for Materials Research, Tohoku University, Sendai 980-8577, Japan \\
$^4$RIKEN Center for Emergent Matter Science (CEMS), Wako 351-0198
}

\date{\today}

\begin{abstract}

We present NMR measurements of the layered nitride superconductor Li$_x$ZrNCl.
The nuclear spin-lattice relaxation rate, $1/T_1$, shows that the coherence peak is strongly suppressed in Li$_x$ZrNCl in contrast to conventional BCS superconductors.
In the lightly-doped region close to the insulating state, the system shows a gap-like behavior, i.e., pseudogap, that is characterized by a reduction in the magnitude of the Knight shift and $1/T_1T$.
A higher superconducting (SC) transition temperature, $T_c$, is achieved by coexisting with the pseudogap state.
These unusual behaviors, which deviate from the ordinary BCS framework, are the key ingredients to understanding the SC mechanism of Li$_x$ZrNCl.

\end{abstract}

\maketitle

Carrier doping into semiconductors and band-insulators can induce superconductivity, as seen in covalent crystals such as diamond and silicon \cite{Ekimov,Bustarret}.
In most cases, superconductivity in such systems is understood in the framework of the conventional electron-phonon coupling mechanism based on BCS theory \cite{C_isotope}.
In some exceptional systems, however, a deviation from the ordinary phonon-mediated mechanism is discussed, and the intercalated layered nitride $A$NCl ($A=$ Hf, Zr, or Ti) is one such fascinating example \cite{Yamanaka1,Yamanaka2,Zhang}.
The critical temperature $T_c$ is as high as $\sim26$ K in Hf systems, whereas the electron-phonon coupling is weak, and the density of states at the Fermi level, $D(E_F)$, is low to account for the high $T_c$ \cite{Tou,Taguchi_C}. 
In addition, the isotope effect of N is quite small \cite{Tou_NMR,Taguchi_isotope}.
The higher $T_c$ obtained by the two-dimensional separation of the conducting layer implies the presence of a mechanism beyond that of the ordinary BCS \cite{Takano}.
A key feature is an increase in $T_c$ toward the insulating state, as observed in Li$_x$ZrNCl \cite{Taguchi_doping}.
Systematic studies with Li$_x$ZrNCl have revealed that the size and anisotropy of the superconducting (SC) gap depend on the doping level \cite{Kasahara,Hiraishi}.
These features cannot be explained within the ordinary BCS framework.
Recent calculations based on density functional theory by Akashi {\it et al.} have shown that the high $T_c$ and its doping dependence cannot be reproduced by the ordinary BCS mechanism \cite{Akashi}, whereas Yin {\it et al.} have pointed out that the high $T_c$ can be explained by considering long-range exchange interaction \cite{Yin}.
Experimental confirmation is still insufficient to reach a consensus on the SC mechanism in layered nitride superconductors.

NMR is a powerful tool to investigate the SC order parameter and underlying correlations.
Thus far, few NMR results have been reported for the layered nitride superconductors using Li$_{0.5}$(THF)$_y$HfNCl \cite{Tou_NMR,Tou_int,Hotehama} and Li$_x$ZrNCl \cite{Tou_PhysicaC}.
Spin-singlet pairing has been confirmed from the reduction of spin susceptibility observed via the Knight shift \cite{Tou_NMR,Tou_int,Tou_PhysicaC}.
On the other hand, the nuclear spin-lattice relaxation rate ($1/T_1$) suffers a large contribution from the vortex dynamics induced in the two-dimensional layered structure where molecules are co-intercalated with alkali-ions between the conducting layers \cite{Hotehama}.
Unfortunately, this contribution has prevented the evaluation of intrinsic relaxation by quasiparticles; therefore, a conclusion about the SC symmetry has not been attained \cite{Tou_int,Hotehama}.
In this Rapid Communication, we report NMR results of Li$_x$ZrNCl where the contribution from the vortex dynamics inherent in the two-dimensional superconductor is successfully excluded in the relaxation process because of the short distance between the conducting layers.
The $1/T_1$ suggests that strong suppression of the coherence effect is an intrinsic property of this material, and $T_c$ is enhanced by coexisting with a pseudogap state, which emerges from a temperature higher than $T_c$ in the lightly-doped region.

The $^{15}$N isotope-enriched polycrystalline samples were prepared using the same procedure as that given in Ref.~12, where a high degree of $c$-axis orientation is obtained by compressing the powder sample.
The distribution of $c$-axis orientation is checked to be within $\pm5$ \% from X-ray rocking curve experiment.
The samples were sealed inside a quartz tube to avoid oxidization, and NMR measurements were performed using the single-pulse method for the $^{15}$N nucleus and the spin-echo method for the $^{91}$Zr nucleus.

\begin{figure}[htb]
\centering
\includegraphics[width=0.8\linewidth]{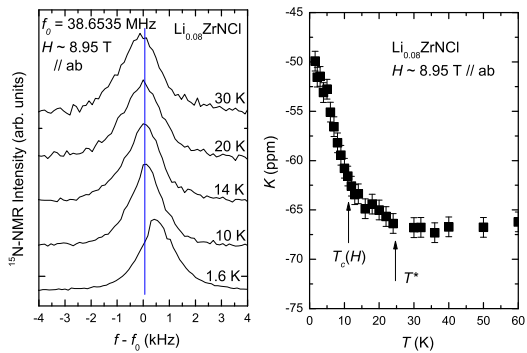}
\caption[]{(color online) (a) $^{15}$N-NMR spectrum of Li$_x$ZrNCl ($-1/2 \leftrightarrow 1/2$ transition) for $H \parallel ab$ for $x=0.08$. The guideline is drawn at the central position of the spectrum at 14 K just above $T_c(H)$. (b) Temperature dependence of the Knight shift, $K$. $K$ increases below $T_c(H)$ owing to spin-singlet pairing, and a gradual variation of $K$ is observed below $T^*$.
}
\end{figure}

Figure 1(a) shows $^{15}$N-NMR spectra of Li$_x$ZrNCl measured with a magnetic field of $H\sim8.95$ T parallel to the $ab$ plane for $x=0.08$.
A clear shift is observed at low temperatures owing to the occurrence of superconductivity at $T_c(H)\sim11$ K, whereas the shift is present in the high temperature range above $T_c$.
Figure 1(b) shows the temperature dependence of the Knight shift ($K$), which is estimated from the isotropic part of the chemical shift for pristine ZrNCl \cite{Tou_PhysicaC}.
The $K$ begins to increase from $T^* \sim 25$ K and exhibits a steep increase below $T_c(H)$.
The increase in $K$ below $T_c(H)$ clearly indicates a decrease in spin susceptibility through a negative hyperfine coupling constant \cite{Tou_NMR,Tou_PhysicaC}, ensuring that the superconductivity is of bulk nature even for $x=0.08$, close to the insulating state.
The anomaly below $T^*$ indicates that the spin susceptibility decreases from a temperature higher than $T_c$.
The Knight shift for $x=0.08$ has been reported using a batch different from the present sample under $H\sim 4$ T \cite{Tou_PhysicaC}, where a gradual shift was present in the temperature range above $T_c$, although large errors originating from measurements under a lower field makes the anomaly unclear.
Considering the diamagnetic shift of $\sim15$ ppm estimated from the shift at the Cl site, the reduction in $K$ in the SC state is evaluated to be $\sim30$ ppm, which is comparable to that estimated under $H \sim4$ T \cite{Tou_PhysicaC}.
Thus, the magnitude of the spin part of $K$ in the normal state is roughly estimated to be $\sim30$ ppm for $H \parallel ab$.

\begin{figure}[htb]
\centering
\includegraphics[width=0.65\linewidth]{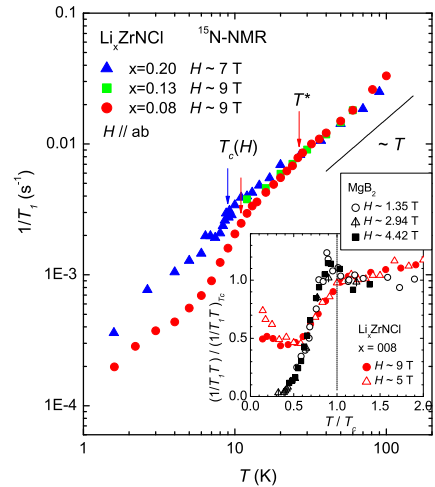}
\caption[]{(color online) Temperature dependences of $^{15}$N-$1/T_1$ in Li$_x$ZrNCl for $x=0.08, 0.13$, and $0.20$. $1/T_1$ is independent of $x$ in the normal state above $T^*$. A clear decrease below $T_c(H)$ appears in $1/T_1$ without the coherence peak for $x=0.08$. The inset shows a comparison between MgB$_2$  and Li$_x$ZrNCl. 
}
\end{figure}

Figure 2 shows the temperature dependences of $^{15}$N-$1/T_1$ in Li$_x$ZrNCl for different doping levels.
In the normal state, $1/T_1$ follow the relationship $T_1T=const.$, which is a usual metallic behavior, except for $x=0.08$.
The $1/T_1$ for $x=0.08$ decreases more steeply than the slope proportional to $T$ below $T^*$.
The values of $1/T_1$ above $T^*$ are almost independent of the doping level.
In the SC state, $1/T_1$ for $x=0.08$ shows a clear decrease below $T_c(H)$ without any signature of the coherence peak (Hebel-Slichter peak), which appears in ordinary BCS $s$-wave superconductors \cite{Hebel}.
The inset of Fig.~2 shows a comparison between Li$_{0.08}$ZrNCl and a phonon-mediated superconductor MgB$_2$ \cite{Kotegawa_MgB2}.
The observed trend for $1/T_1T$ in MgB$_2$ shows the coherence peak just below $T_c$, and decreases markedly below $\sim0.8 T_c$, in sharp contrast to $1/T_1T$ in Li$_{0.08}$ZrNCl.
In general, the coherence peak is suppressed in high-$T_c$ strong electron-phonon coupling superconductors, including MgB$_2$, owing to the short lifetime of the quasiparticles; however, complete suppression of the coherence peak is not attained even in MgB$_2$.
Moreover, a magnetic field suppresses the coherence peak, but the coherence peak of MgB$_2$ is robust against a magnetic field of $\sim4.4$ T, which is about 30\% of the critical field $H_{c2}\sim16$ T \cite{Angst,Lyard}.
In Li$_{0.08}$ZrNCl, on the other hand, the coherence peak does not recover at all, even under a magnetic field of 5 T,  which is sufficiently low compared with the critical field, $H_{c2}\sim20$ T for $H \parallel ab$ \cite{Tou_Hc2}.
Thus, the magnetic field is not a main factor for the suppression of the coherence peak in Li$_{0.08}$ZrNCl.
Specific heat measurements suggest that a clear isotropic gap opens in the lightly-doped region \cite{Kasahara}. 
Nevertheless, the coherence peak is strongly suppressed in Li$_{0.08}$ZrNCl.
Therefore, the strong suppression should be considered as another intrinsic feature related to the SC mechanism in this system.

$1/T_1T$ has a substantial value at low temperatures, but this is likely to be owing to a contribution other than the quasiparticles, because it exhibits a field dependence between 5 and 9 T.
We consider that the extrinsic contribution, which is conjectured to be of the same origin as that causing Curie behavior in the bulk susceptibility \cite{Kasahara}, dominates the relaxation process at the N site at low temperatures because of the long $T_1$ ($\sim5000$ s at 1.6 K).

\begin{figure}[htb]
\centering
\includegraphics[width=0.6\linewidth]{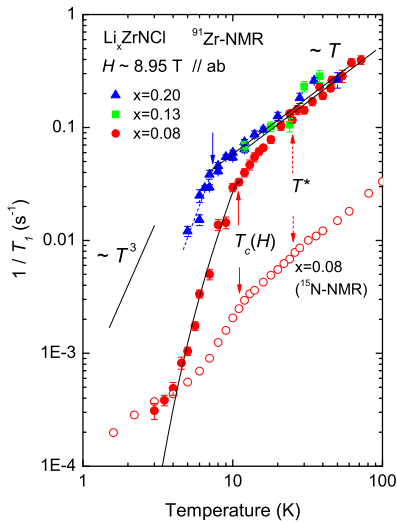}
\caption[]{(color online) Temperature dependences of $^{91}$Zr-$1/T_1$ in Li$_x$ZrNCl for $x=0.08, 0.13$, and $0.20$. $1/T_1$ at the Zr site is 10 or more times larger than that at the N site. The solid curve is a calculated result assuming an isotropic gap of $2\Delta/k_BT_c = 4.5$ and the absence of the coherence effect.

}
\end{figure}

Figure 3 shows the temperature dependences of $^{91}$Zr-$1/T_1$ in Li$_x$ZrNCl for $x=0.08, 0.13$, and $0.20$.
The anomaly below $T^*$ was also observed at the Zr site for $x=0.08$.
$1/T_1$ for $T>T^*$ is almost independent of $x$ as was the case of the N site.
$1/T_1T$ at the Zr site is 10 or more times larger than that at the N site because the partial $D(E_F)$ is larger at the Zr site \cite{Weht,Hase,Felser,Akashi}, in addition to the difference in the hyperfine coupling constants.
For $x=0.08$ and $x=0.20$, $1/T_1$ shows a steep decrease below $T_c(H)$.
Because of the shorter $T_1$ than at the N site, $1/T_1$ at the Zr site can avoid the relaxation from the extrinsic origin, even below $T_c(H)$, and $1/T_1$ for $x=0.08$ shows a large decrease and saturates at a similar value to that at the N site.
For $x=0.20$, we omitted the data below $\sim5$ K because the emergence of an unexpected long component in $T_1$ disturbed the determination of a unique and reliable $T_1$.
An obvious coherence peak was not observed even at the Zr site for $x=0.08$ and $x=0.20$.
The temperature dependence of $1/T_1$ for $x=0.08$ is steeper than $T^3$ in the SC state.
The solid curve in the figure is $1/T_1$ calculated using an isotropic SC gap model with the complete exclusion of the coherence effect, where the size of the SC gap is $2\Delta/k_BT_c = 4.5$.
This gap size is in good agreement with the specific heat \cite{Kasahara}.

\begin{figure}[htb]
\centering
\includegraphics[width=0.7\linewidth]{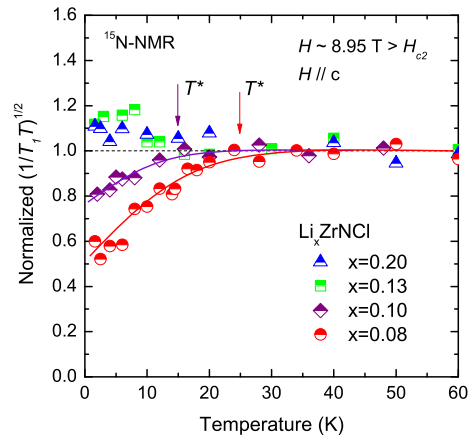}
\caption[]{(color online) Temperature dependences of $[(T_1T)_n/T_1T]^{1/2}$ measured for $H \parallel c$, where $(T_1T)_n$ is the value above $T^*$. Superconductivity is suppressed by the magnetic field of $\sim9$ T.  The gradual decrease below $T^*$ is robust against the magnetic field comparable to $H_{c2}$. The solid curves serve as a visual guide.
}
\end{figure}

Figure 4 shows the normalized $(1/T_1T)^{1/2}$, which is proportional to the spin susceptibility in ordinary metals, measured at the N site under $H \parallel c$.
Li$_x$ZrNCl has a smaller $H_{c2}$ for $H \parallel c$ compared with $H \parallel ab$, and a magnetic field of $H \sim 9$ T ($\parallel c$) almost suppresses the SC state \cite{Takano_PRB}.
$(1/T_1T)^{1/2}$ under $H \parallel c$ also shows the reduction below $T^*$ for $x=0.08$ and $x=0.10$, indicating that this behavior is robust against the magnetic field comparable to $H_{c2}$.
The energy scale of this gap-like behavior is much lower than the band gap of the pristine ZrNCl ($\sim2-3$ eV) \cite{Yamanaka1,Weht,Hase,Felser,Akashi}.
Therefore, this gap-like behavior is denoted as a "pseudogap" behavior.

\begin{figure}[b]
\centering
\includegraphics[width=0.8\linewidth]{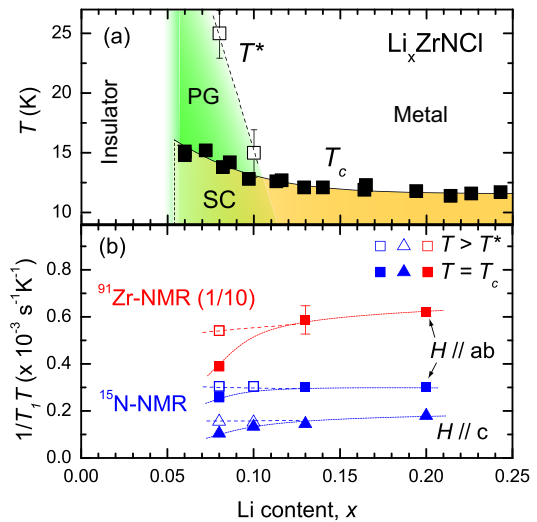}
\caption[]{(color online) The doping level $x$ dependences of (a) $T_c$, $T^*$, and (b) $1/T_1T$ above $T^*$ and at $T_c(H=0)$. The values of $T_c$ were obtained from \cite{Taguchi_doping}. A pseudogap behavior (PG in the figure) appears near the insulating state, and $T_c$ is enhanced in the pseudogap state. The anisotropy of $^{15}$N-$1/T_1T$ originates in the anisotropy of the hyperfine coupling constant.
}
\end{figure}

Figure 5(a) shows a phase diagram of Li$_x$ZrNCl.
The pseudogap state detected by the Knight shift and $1/T_1T$ appears near the insulating state.
$T_c$ is almost unchanged at the higher doping level, whereas it increases towards the insulating state accompanied by the appearance of the pseudogap state.
This suggests that the electronic state in the pseudogap state is a key factor for enhancement of $T_c$ in the lightly-doped region.
In Fig.~5(b), we show the doping level dependence of $1/T_1T$ above $T^*$ and at $T_c(H=0)$ for both N and Zr sites.
$1/T_1T$ above $T^*$ is almost independent of $x$ for both sites, and $1/T_1T$ at $T_c$ is suppressed near the insulating state because of the development of the pseudogap state.
It is obvious that $T_c$ is inversely correlated with $1/T_1T$, which is proportional to $\alpha D^2(E_F)$, where $\alpha$ is a factor representing electronic correlations \cite{Moriya}.
The inverse correlation between $1/T_1T$ and $T_c$ cannot be explained in the ordinary BCS framework.
The absence of the coherence peak also suggests that the SC mechanism in layered nitride superconductors is beyond the simple BCS framework, at least in the lightly-doped region.

It is important to compare the pseudogap behavior with other measurements in similar temperature and doping ranges. In such ranges, the resistivity shows a remarkable semiconducting behavior and the magnitude of the Hall coefficient increases toward low temperatures \cite{Takano_PRB}. The bulk susceptibility shows a steep decrease below $\sim50$ K for $x=0.08$ \cite{Kasahara}, but it cannot be determined whether this decrease corresponds to the reduction in $1/T_1T$ because the gradual decrease in the bulk susceptibility remains in the highly-doped region. As for the specific heat, the temperature dependence has been reported as $\Delta C(H,T) = C(H,T) - C(H>H_{C2},T)$ \cite{Kasahara}. Therefore, if the anomaly is insensitive to a magnetic field, it is canceled out in $\Delta C(H,T)$.
On the other hand, the $x$-dependence of the normal-state Sommerfeld constant $\gamma_n$ has also been estimated from a recovery of the specific heat under a magnetic field \cite{Kasahara}.
The value of $\gamma_n$ was found to decrease gradually toward the insulating state, but its reduction is likely to be weaker than $1/T_1T$, even though $1/T_1T$ generally corresponds to $\gamma_n^2$.
The reduction in $1/T_1T$ might be enhanced by the change in magnetic correlations and the hyperfine coupling constant.

We discuss the cause of the enhancement of $T_c$ toward the insulating state, which must be related with the SC mechanism of Li$_x$ZrNCl.
The first candidate is a nesting-induced spin-fluctuation scenario \cite{Kuroki}.
This gives the SC gap of a $d+id'$ symmetry without nodes, which can account for the unusual doping dependences of $T_c$ \cite{Kasahara}.
The isotropic gap \cite{Kasahara} and absence of the coherence peak in $1/T_1$ also can be explained by this gap function.
The calculated susceptibility is temperature independent, which has no discrepancy in the $T_1T \sim constant$ behavior observed over a wide doping region \cite{Kuroki}.
On the other hand, the calculated susceptibility is enhanced toward the lightly-doped region because of its better nesting property \cite{Kasahara}.
This differs from the doping dependence of $1/T_1T$.
In our measurements, there was no signature indicating that spin fluctuations develop with decreasing $x$ accompanied by the increase in $T_c$.
The similarity of behavior for $1/T_1T$ between the N and Zr sites rules out the possibility that magnetic fluctuations are accidentally canceled by the form factor at each site.
Since $1/T_1T$ is a low-energy probe, the remaining issue is the possibility that high-energy spin fluctuation is a crucial role for superconductivity.
If this is correct, the pseudogap in $1/T_1T$ might include a spin gap in which  the spectral weight of the spin fluctuations transfers to the high-energy region.
Theoretical and experimental investigations for the energy dependence of the spin fluctuations are required to confirm this.

The second candidate is a scenario whereby $T_c$ is enhanced by Anderson localization in the lightly-doped region \cite{Feigelman,Yanase,Burmistrov,Kravtsov}.
It has been pointed out that the fractal (inhomogeneous) wave function affected by randomness near the Anderson transition can enhance the pairing interaction on the site contributing to conductivity in the $s$-wave framework. If the coherency of the pairing develops, a higher $T_c$ is realized. 
It seems likely that the pseudogap in $1/T_1T$ near the insulating state originates from the localization of carriers at the impurity level at low temperatures.
This is consistent with Hall effect measurements, where the Hall coefficient shows a distinct temperature dependence for $x=0.07$ \cite{Takano_PRB}.
If the system approaches the Anderson localization, the transferred hyperfine coupling constant from neighboring sites is expected to decrease. This might give a strong reduction in $1/T_1T$ below $T^*$.
Interestingly, the phase diagram shown in Fig.~5(a), resembles the theoretical predictions well, where $T_c$ increases toward the Anderson localization \cite{Feigelman,Yanase,Burmistrov,Kravtsov}.
In these models, the pseudogap owing to the short-range incoherent Cooper pairs has been predicted \cite{Feigelman,Yanase}.
It is an intriguing issue that the pseudogap observed in $1/T_1T$ and the Knight shift includes the formation of incoherent Cooper pairs.
Such short-range Cooper pairs have been expected to be more robust against an external magnetic field than the coherent superconductivity \cite{Feigelman}.
In addition, the coherence peak in $1/T_1$ is expected to be suppressed in the localization regime because strong scattering is present on account of the disorder.
In this scenario, the unsolved issue is whether the relatively high $T_c$ in the higher doping region can be explained by electron-phonon coupling.
This may be reconciled in terms of either underestimation of the electron-phonon coupling \cite{Yin} or the presence of an additional contribution such as that of a plasmon \cite{Bill1,Bill2,Akashi_plasmon}.

In summary, we performed $^{15}$N- and $^{91}$Zr-NMR measurements on the layered nitride superconductor Li$_x$ZrNCl.
The coherence peak in $1/T_1$ was strongly suppressed, which is in sharp contrast to conventional BCS superconductors, including the strong coupling MgB$_2$.
The pseudogap state, characterized by the reduction of $1/T_1T$ and the Knight shift, appears in the lightly-doped region.
A higher $T_c$ is realized in the pseudogap state, where the Fermi liquid state breaks down.
These key ingredients will help us understand the SC mechanism of Li$_x$ZrNCl that lies beyond ordinary BCS theory.

\section*{Acknowledgements}

We thank  K. Kuroki, Y. Yanase, and R. Arita for their valuable discussions.
This work has been partly supported by Grants-in-Aid for Scientific Research (Nos. 24340085, 19204036, 20045010, and 22013011) from the Ministry of Education, Culture, Sports, Science and Technology (MEXT) of Japan.

\end{document}